\documentclass[reprint,superscriptaddress, amsmath,amssymb, prl]{revtex4-2}

\usepackage{graphicx}
\usepackage{subfigure}
\usepackage{dcolumn}
\usepackage{bm}
\usepackage[colorlinks, linkcolor=red, anchorcolor=blue, citecolor=blue,urlcolor=black]{hyperref}
\usepackage{ulem}
\usepackage{verbatim}
\usepackage{mathrsfs}

\newcommand{\beq}{\begin{eqnarray} }
\newcommand{\eeq}{\end{eqnarray} }
\newcommand{\Beq}{\begin{eqnarray*} }
\newcommand{\Eeq}{\end{eqnarray*} }
\newcommand{\Bmat}{\left(\begin{matrix}}
\newcommand{\Emat}{\end{matrix}\right)}

\begin{document}


\title{Large intrinsic anomalous Hall effect in both Nb$_{2}$FeB$_{2}$ and Ta$_{2}$FeB$_{2}$ with collinear antiferromagnetism}

\author{Xiao-Yao Hou}
\affiliation{Department of Physics and Beijing Key Laboratory of Opto-electronic Functional Materials $\&$ Micro-nano Devices, Renmin University of China, Beijing 100872, China}
\author{Huan-Cheng Yang}
\affiliation{Department of Physics and Beijing Key Laboratory of Opto-electronic Functional Materials $\&$ Micro-nano Devices, Renmin University of China, Beijing 100872, China}
\author{Zheng-Xin Liu}
\affiliation{Department of Physics and Beijing Key Laboratory of Opto-electronic Functional Materials $\&$ Micro-nano Devices, Renmin University of China, Beijing 100872, China}
\author{Peng-Jie Guo}\email{guopengjie@ruc.edu.cn}
\affiliation{Department of Physics and Beijing Key Laboratory of Opto-electronic Functional Materials $\&$ Micro-nano Devices, Renmin University of China, Beijing 100872, China}
\author{Zhong-Yi Lu}\email{zlu@ruc.edu.cn}
\affiliation{Department of Physics and Beijing Key Laboratory of Opto-electronic Functional Materials $\&$ Micro-nano Devices, Renmin University of China, Beijing 100872, China}
\noaffiliation

\date{\today}

\begin{abstract}

It is rarely reported that collinear antiferromagnetic (AFM) metals can have anomalous Hall effect (AHE). In this letter, based on symmetry analysis and the first-principles electronic structure calculations, we predict that two existing collinear antiferromagnets Nb$_{2}$FeB$_{2}$ and Ta$_{2}$FeB$_{2}$, whose Néel temperatures are above room temperature, have very large AHE with anomalous Hall conductance (AHC) -100 $\Omega^{-1}$ cm$^{-1}$ and $-54\Omega^{-1}$ cm$^{-1}$, respectively. We further complete the symmetry resquirements for realizing the AHE in collinear antiferromagnetism. 

\end{abstract}

\maketitle
{\it Introduction.} 
Anomalous Hall effects are usually observed in ferromagnetic metals, and the corresponding Hall conductance is proportional to the magnetization. The AHE generally contains the contributions from both extrinsic and intrinsic mechanisms\cite{RMP-2010, RMP-2015}.
The intrinsic contribution of AHC comes from the Berry curvature in the Brillouin zone below the Fermi surface. Just like external magnetic field, the Berry curvature varies like a (pseudo) vector representation under the action of symmetry operation. Thus the symmetry group determines the existence of intrinsic AHE effect. For example, if time-reversal ($\mathcal T$) is a symmetry, then the integration of Berry curvature will be precisely zero. This indicates that the AHE can’t exist in non-magnetic metals. In contrast, for a antiferromagnet, if its magnetic order breaks all the symmetries of its crystal, then the AHC may be non-zero. For instance, a noncollinear antiferromagnetic (AFM) order generally breaks all the crystal point group symmetries, therefore the realization of AHE is relatively natural. 
The first predicted noncollinear AFM metal as an AHE candidate is Mn$_{3}$Ir\cite{Chen-prl, Epl-2014}. 
Later, large AHC has been experimentally observed in non-collinear antiferromagnets Mn$_{3}$Sn\cite{Nature-2015, AM-2019}, Mn$_{3}$Ge\cite{PRA-2016, SA-2016, ACS-2020}, Mn$_{3}$Pt\cite{NE-2018, PRM-2021}.

For collinear antiferromagnetism, Libor Šmejkal et al predicted that the collinear AFM metals RuO$_{2}$ and CoNb$_{3}$S$_{6}$ have non-zero AHC\cite{CAHE-2020}. However, since the intrinsic easy magnetization axis of RuO$_{2}$ is along the $\left[ 001 \right]$ rather than the $\left[ 100 \right]$ or $\left[ 110 \right]$ directions, the AHC of RuO$_{2}$ is eventually zero owing to two mirror symmetries perpendicular to each other ($M_{z}$ and $\{ M_{x} | (1/2, 1/2, 1/2)\}$)\cite{Feng-2020}. The other AHE candidate CoNb$_{3}$S$_{6}$, unfortunately, was also excluded because the AFM order was shown to be noncollinear rather than collinear\cite{NC-2018, PRM-2022}. In comparison with noncollinear AFM, collinear AFM orders have relatively high symmetries. Especially, the so-called ‘prohibiting’ symmetry $\{\mathcal T| \tau\}$ or $\mathcal I \mathcal T$ (the $\tau$ denotes the associated fractional translation, and $\mathcal I$ represents the spatial inversion) usually exists in collinear AFM systems, which restricts the AHC to be zero\cite{Guo-2022}. Fortunately, type-I and part of type-III magnetic space groups do not contain the ‘prohibiting’ symmetry elements, which makes it possible to realize AHE and even quantum anomalous Hall effect (QAHE) in collinear antiferromagnetic systems\cite{Guo-2022}. Overall, to realize the AHE in collinear AFM, one needs to overcome three challenges: i) the AFM materials are generally insulators rather than metals; ii) the collinear AFM materials have usually the ‘prohibiting’ symmetry $\{\mathcal T| \tau\}$ or $\mathcal I \mathcal T$; iii) for high-symmetry crystal, there are likely additional symmetry constraints restricting the AHC to be zero, for example RuO$_{2}$. Considering that the temperature for observing the AHE may be substantially high since the Néel temperature of the AFM order can be very high, searching for collinear AFM metals with large AHC is thus extremely important in this field.

In this letter, based on symmetry analysis and the first-principles electronic structure calculations, we predict two AHE candidates with collinear AFM order. We firstly demonstrate that the magnetic orders of Nb$_{2}$FeB$_{2}$ and Ta$_{2}$FeB$_{2}$ both are collinear AFM and have no ‘prohibiting’ symmetry. Secondly, the calculated Néel temperatures of Nb$_{2}$FeB$_{2}$ and Ta$_{2}$FeB$_{2}$ are 433 K and 401 K, respectively. Finally, we show that both Nb$_{2}$FeB$_{2}$ and Ta$_{2}$FeB$_{2}$ have large AHC. For instance, the AHC of Nb$_{2}$FeB$_{2}$ is of order of -100 $\Omega^{-1}$ cm$^{-1}$, which is the same order of magnitude as those of ferromagnetic metals. Therefore, these two collinear AFM metals may exhibit large AHE at room temperatures.

 \begin{figure*}
\centering
\includegraphics[width=15cm]{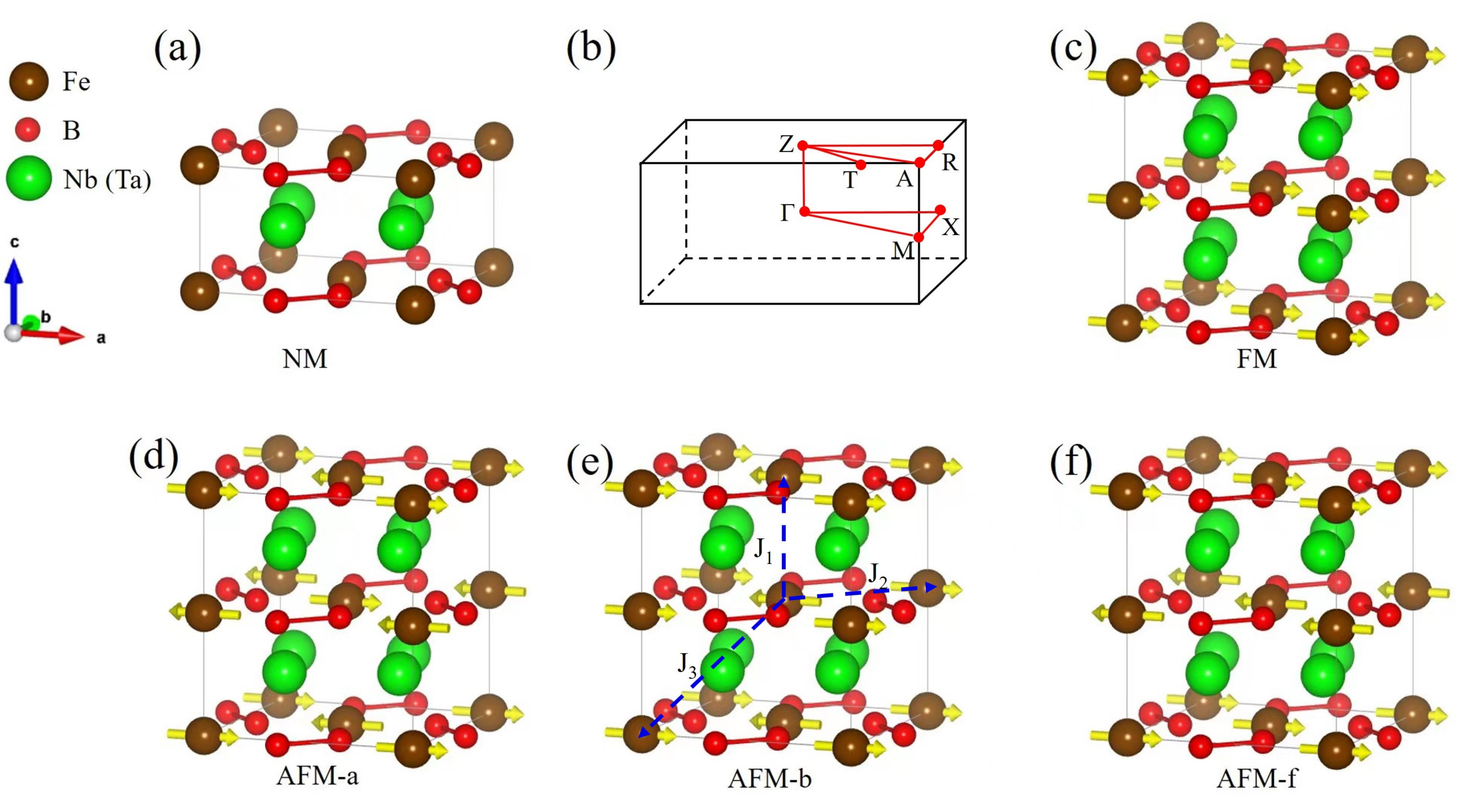}
	\caption{(a) The nonmagnetic primitive cell and (b) Bulk Brillouin zone (BZ) of Nb$_{2}$FeB$_{2}$ and Ta$_{2}$FeB$_{2}$. The high symmetry points and paths are marked by red dots and lines. Four typical collinear magnetic configurations for Nb$_{2}$FeB$_{2}$ and Ta$_{2}$FeB$_{2}$: (c) FM state, (d) the AFM state with both intralayer and interlayer AFM couplings, (e) the AFM state with intralayer AFM and interlayer FM couplings, (f) the AFM state with intralayer FM and interlayer AFM couplings. The first-, second- and third-nearest neighbours are labeled in the figure(e). Here yellow arrows represent spins.}\label{fig:1}
\end{figure*} 

 {\it Method.} The first-principles electronic structure calculations were performed in the framework of density functional theory (DFT)\cite{PR-1964, PR-1965} using the Vienna Ab-initio Simulation Package (VASP)\cite{CMS-1996, PRB-1993, PRB-1996}. The generalized gradient approximation (GGA) of Perdew-Burke-Ernzerhof (PBE) type\cite{GGA-PRL} was adopted for the exchange-correlation functional. The projector augmented wave (PAW) method\cite{PRB-1999} was adopted to describe the interactions between valence electrons and nuclei. The kinetic energy cutoff of the plane-wave basis was set to be 600 eV. The total energy convergence and the atomic force tolerance were set to be 10$^{-6}$ eV and 0.01 eV/\AA, respectively. For describing the Fermi-Dirac distribution function, a Gaussian smearing of 0.05 eV was used. The $8\times 8 \times 15$ and $9\times 9 \times 8$ Monkhorst-Pack $k$ meshes were used for the Brillouin zone sampling of the unit cell and the $1 \times 1\times 2$ supercell, respectively. To account for the correlation effects of Fe 3d orbitals, we performed GGA+U\cite{LDAU-1991} calculations by using the simplified rotationally invariant version of Dudarev et al\cite{PRB-1998}. The onsite effective U$_{\rm eff}$ values of Fe-d electrons in Nb$_{2}$FeB$_{2}$ and Ta$_{2}$FeB$_{2}$ were estimated by the linear response method\cite{PRB-2005}. The Berry curvature and AHC were calculated by using the open source Wannier90\cite{wann-1997, wann-2001}. The Monte Carlo simulations based on the classical Heisenberg model were performed by using the open source project MCSOLVER\cite{ASS-2019}. 

{\it Results.} The transition metal borides Nb$_{2}$FeB$_{2}$ and Ta$_{2}$FeB$_{2}$ crystalize in the U$_{3}$Si$_{2}$ tetragonal structure type with space group P4-mbm (No. 127). The elementary symmetry operations of the P4-mbm space group are $C_{4z}$, $\{ C_{2x} | (1/2, 1/2)\}$, and $ \mathcal I$, yielding the point group $D_{4h}$. A schematic diagram of the unit cell is shown in Fig.\ref{fig:1}(a), which adopts a layer-like structure and is built up by FeNb (FeTa) units with CsCl structure and NbB$_{2}$ (TaB$_{2}$) units with face-connected AlB$_{2}$ structure motif. The corresponding BZ together with the high-symmetry points and high symmetry lines are shown in Fig.\ref{fig:1}(b).

\begin{figure}
	\includegraphics[width=7.5 cm]{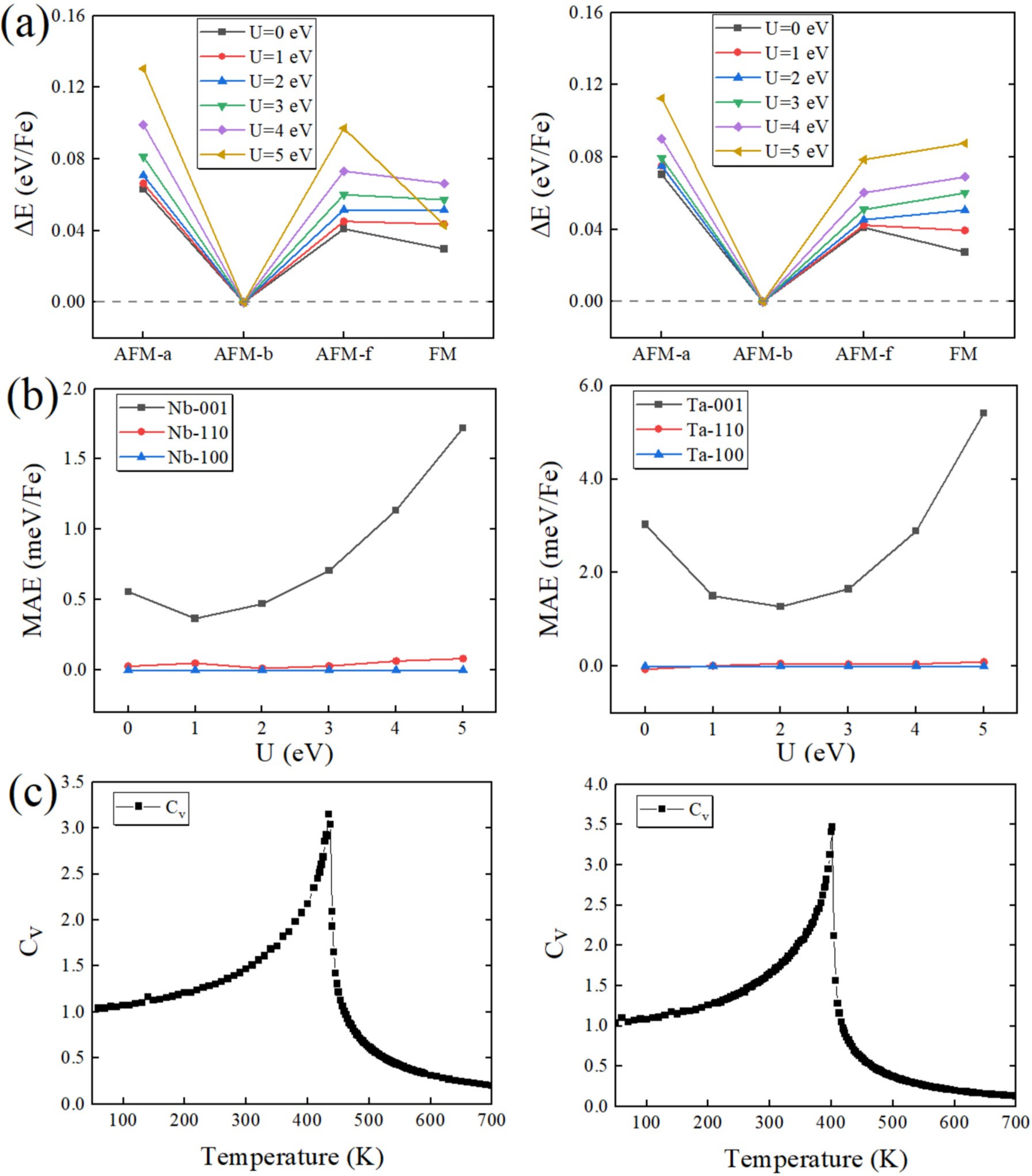}
	\caption{(a) the relative energies of three AFM states with respect to the FM state, (b)the magnetic anisotropy energy (MAE) of the ground state (AFM-b), (c) the evolution of specific heat capacity with temperature. Left panels and right panels are for Nb$_{2}$FeB$_{2}$ and Ta$_{2}$FeB$_{2}$, respectively.}\label{fig:2}
\end{figure}

In order to determine the ground states of Nb$_{2}$FeB$_{2}$ and Ta$_{2}$FeB$_{2}$, we considered four trial magnetic configurations, one FM state and three possible AFM states (AFM-a, AFM-b and AFM-f) as shown in [Fig.\ref{fig:1}(c)-(f)] within a $1 \times 1\times 2$ supercell. The calculated results show that the AFM-b state is lowest in energy for both Nb$_{2}$FeB$_{2}$ and Ta$_{2}$FeB$_{2}$ (Fig.\ref{fig:2}(a)), which is consistent with the previous calculations\cite{JSSC-2014, CMS-2015}.  The magnetic anisotropy calculations further indicate that the intrinsic easy magnetization axes of both Nb$_{2}$FeB$_{2}$ and Ta$_{2}$FeB$_{2}$ are lying in the $xy$ plane (see Fig.\ref{fig:2}(b)).  We also examined the impact of the onsite Coulomb repulsion U. We find that the AFM ground state is robust for a large range of the values of U (U =0.0, 1.0, 2.0, 3.0, 4.0 and 5.0 eV). In this work we adopted the values of $U_{\rm eff}$ derived from linear response calculations, with  $U_{\rm eff}=$4.82 eV and $U_{\rm eff}$=4.76 eV for Nb$_{2}$FeB$_{2}$  and Ta$_{2}$FeB$_{2}$, respectively. Accordingly, the calculated local moment at each Fe ion is as large as  about 3 $\mu_{B}$ with a quite large magnetic anisotropic energy (MAE) for both Nb$_{2}$FeB$_{2}$  and Ta$_{2}$FeB$_{2}$ (Fig.\ref{fig:2}(b)). Thus the corresponding AFM quantum fluctuation is small.

Then we may estimate the Néel temperatures ($T_{N}$) of Nb$_{2}$FeB$_{2}$ and Ta$_{2}$FeB$_{2}$ from the classical Monte-Carlo simulations based on the 3-dimensional square lattice Heisenberg model  with single-ion anisotropy: 
\begin{eqnarray}
	H = {J_{1}} {\sum_{\langle i, j\rangle} \pmb{S_{i}}\pmb{S_{j}}} 
	+ J_{2}{\sum_{\langle\langle i, j\rangle\rangle} \pmb{S_{i}} \pmb{S_{j}}} \nonumber \\
	+ J_{3}{\sum_{\langle\langle\langle i, j\rangle\rangle\rangle} \pmb{S_{i}} \pmb{S_{j}}}
	 + A \sum_{i}(S_{i}^{x})^{2}, 
	\label{eq:HS}
\end{eqnarray}
where $\pmb{S_{i}}$ represents the spin of Fe ion on the ith site, $J_{1}$, $J_{2}$, and $J_{3}$ denote the first-, second-, and third-nearest exchange interaction parameters, respectively, and $A$ is the single-ion magnetic anisotropy with easy-magnetization axes. The exchange interaction parameters were derived from the mapping analysis of the four magnetic configurations. The results are $J_{1}S^{2}$=-33.78 (-24.69) meV, $J_{2}S^{2}$=6.07 (5.88) meV and $J_{3}S^{2}$ =6.99 (7.03) meV for Nb$_{2}$Fe$B_{2}$ (Ta$_{2}$FeB$_{2}$). The $A( S^{x})^{2}$ deriving from MAE is -1.57 meV (-4.68 meV) for Nb$_{2}$FeB$_{2}$ (Ta$_{2}$FeB$_{2}$). The resultant Néel temperatures for both Nb$_{2}$FeB$_{2}$ and Ta$_{2}$FeB$_{2}$ are 433 K and 401 K, respectively, which are extracted from the peak of the specific heat capacity, as shown in Fig.\ref{fig:2}(c). 

Interestingly, neither $\{\mathcal T| \tau\}$ nor $\mathcal I \mathcal T$ is present in the magnetic structure AFM-b, thus the intrinsic AHE for both Nb$_{2}$FeB$_{2}$ and Ta$_{2}$FeB$_{2}$ may not be prohibited by symmetry. Since the intrinsic easy magnetization axis lies in the $xy$-plane, {\it e.g.} along the $\left[ 100\right]$ direction, 
then the magnetic space group belongs to type-III. The corresponding magnetic point group is $m'm'm$. 
The AHC $\sigma_{xy}$ and $\sigma_{yz}$ must be thus zero according to symmetry constraints but $\sigma_{xz}$ can be nonzero. This is true for both Nb$_{2}$FeB$_{2}$ and Ta$_{2}$FeB$_{2}$ (Fig.\ref{fig:2}(b)). If the intrinsic easy magnetization axis was unfortunately along $\left[ 001\right]$ direction, the magnetic structure would have type-I magnetic space group symmetry, then the mirror symmetries $M_{z}$ and $\{ M_{x} | (1/2, 1/2)\}$ would force the AHC to be zero. 

\begin{figure}
	{\includegraphics[width=7.5 cm]{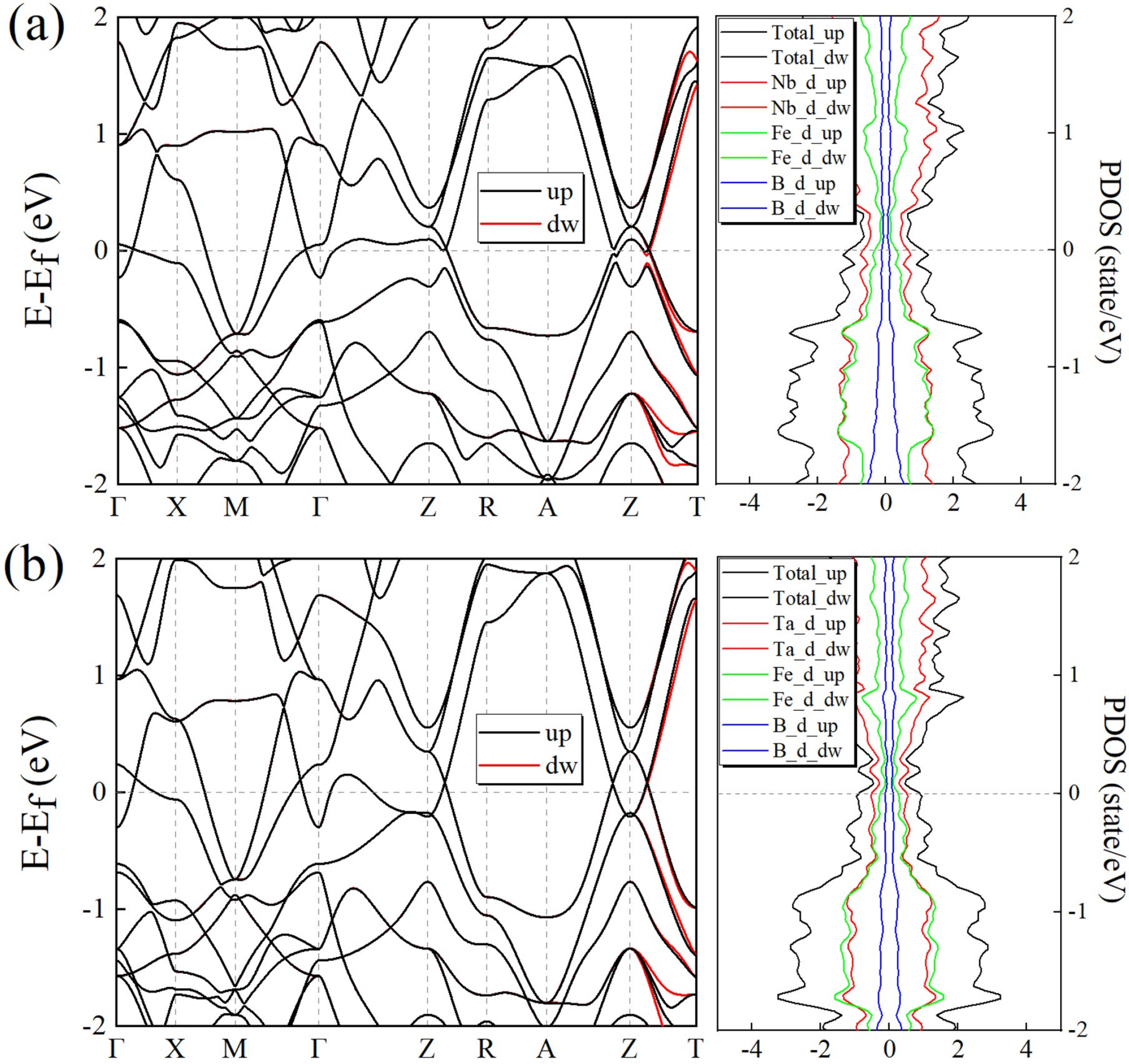}} 
	\caption{The band structure without SOC (left panel) and local density of states (right panel) for Nb$_{2}$FeB$_{2}$ (a) and Ta$_{2}$FeB$_{2}$(b).}\label{fig:3}
\end{figure}

Now we analyze the electronic structures of the collinear AFM Nb$_{2}$FeB$_{2}$ and Ta$_{2}$FeB$_{2}$. When ignoring spin-orbital coupling (SOC), both Nb$_{2}$FeB$_{2}$ and Ta$_{2}$FeB$_{2}$ are collinear AFM metals where the states around the Fermi level are mainly contributed by Nb (Ta) and Fe d orbitals (Fig.\ref{fig:3}(a) and (b)).  
The spin degeneracy is generally lifted due to the breaking of the ‘prohibiting’ symmetries $\{\mathcal T| \tau\}$ and $\mathcal I \mathcal T$. In general $k$ points, such as those in the non-high-symmetry Z-T direction,  the spin degeneracy is obviously split in the Z-T direction (Fig.\ref{fig:3}(a) and (b)). However, the spin degeneracy still remains on all high-symmetry lines, owing to the protection of $\{ C_{2x} \mathcal T | (1/2, 1/2)\}$, $\{ C_{2y} \mathcal T | (1/2, 1/2)\}$, $\{ C_{2xy} \mathcal T | (1/2, 1/2)\}$, and $\{ C_{2x-y} \mathcal T | (1/2, 1/2)\}$ symmetries (see Fig.\ref{fig:3}(a) and (b)). 
Another important feature of the band structures for both Nb$_{2}$FeB$_{2}$ and Ta$_{2}$FeB$_{2}$ is that when ignoring SOC both have Weyl nodal ring around the Fermi level\cite{Next}. 

\begin{figure}
	{\includegraphics[width=5.4 cm]{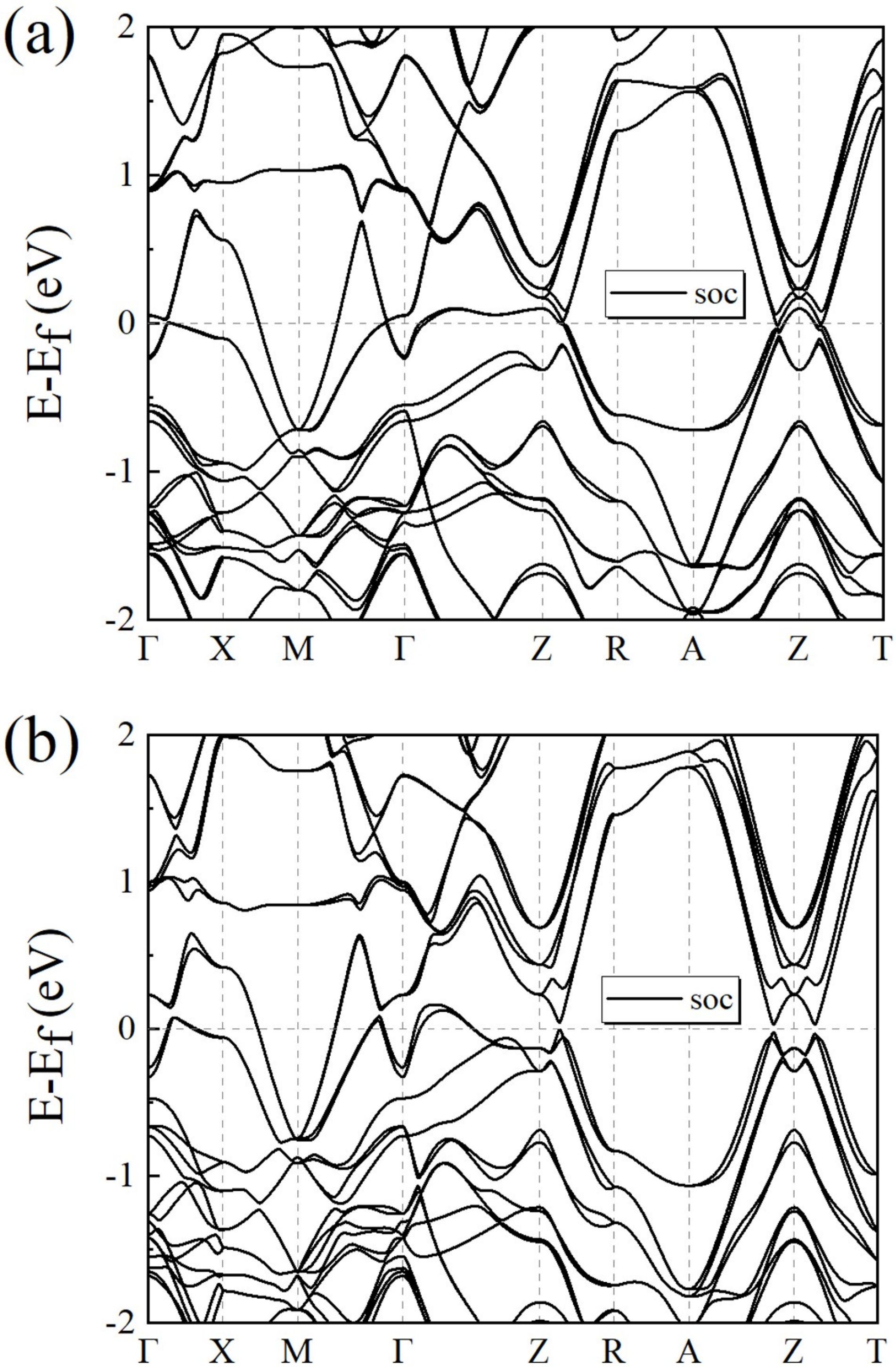}} 
	\caption{The band structure with SOC for Nb$_{2}$FeB$_{2}$ (a) and Ta$_{2}$FeB$_{2}$ (b) with the magnetization along the [100] direction.}\label{fig:4}
\end{figure}

When considering SOC, the intrinsic easy magnetization axis is along the $\left[ 100\right]$ direction and the corresponding symmetry operations are $\{ C_{2y} | (1/2, 1/2)\}$, $\mathcal I$, $\{ M_{y} | (1/2, 1/2)\}$, $C_{2z}\mathcal T$, $M_{z}\mathcal T$, $\{ C_{2x} \mathcal T | (1/2, 1/2)\}$, and $\{ M_{x} \mathcal T | (1/2, 1/2)\}$, therefore the spin degeneracy remains on the high-symmetry lines X-M and R-A directions which are protected by  $\{ C_{2x} \mathcal T | (1/2, 1/2)\}$ and  $\{ M_{x} \mathcal T | (1/2, 1/2)\}$ symmetries (Fig.\ref{fig:4}(a) and (b)).  Meanwhile, the Weyl nodal rings around the Fermi level are lifted by SOC, resulting in large AHC in Nb$_{2}$FeB$_{2}$ and Ta$_{2}$FeB$_{2}$.

\begin{figure}
	{\includegraphics[width=7.5 cm]{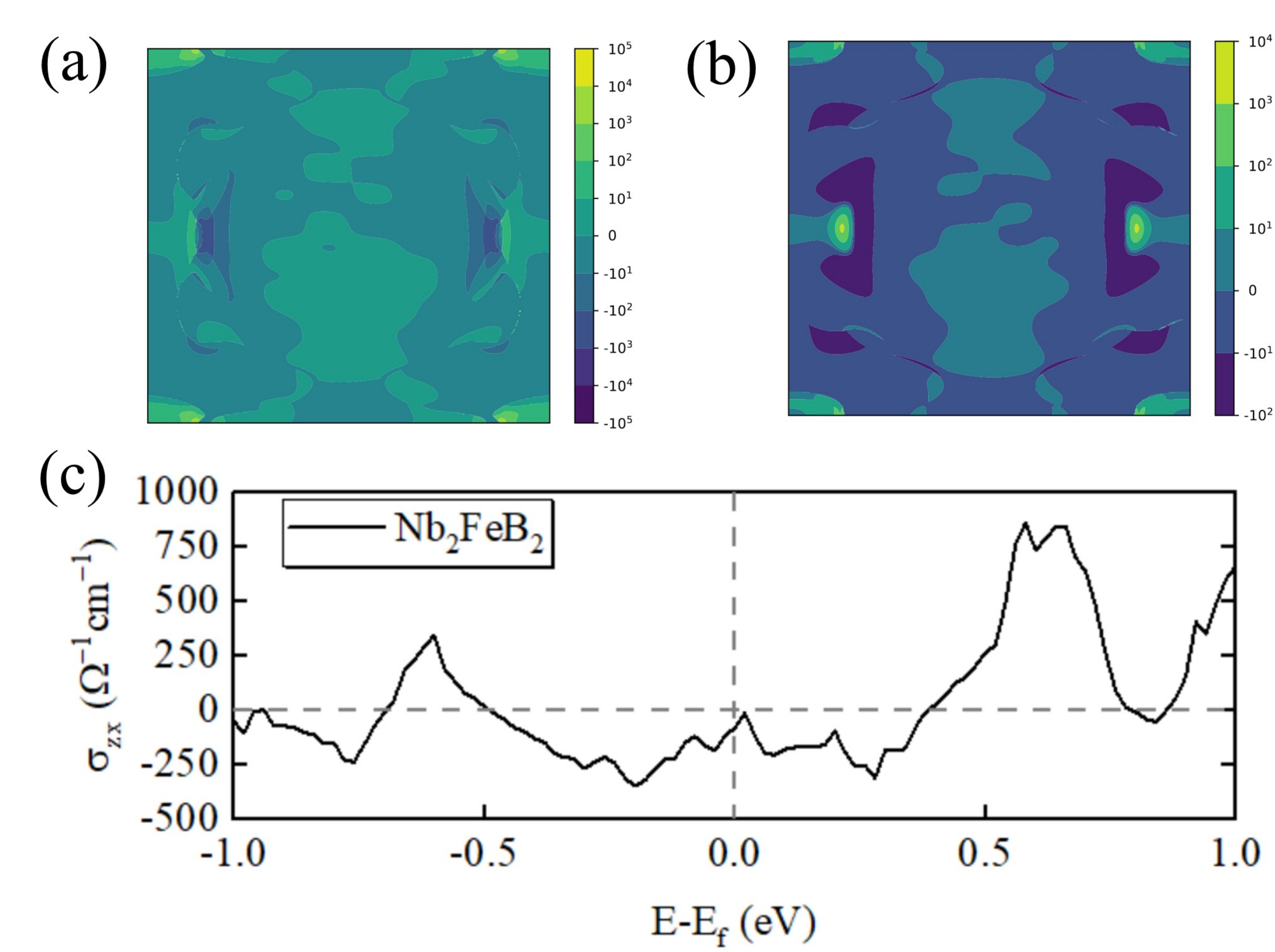}} 
	\caption{The Berry curvature -$\Omega_{y}(k)$ in the $k_{y}$ = 0 plane for Nb$_{2}$FeB$_{2}$ (a) and Ta$_{2}$FeB$_{2}$ (b) at the Fermi level. (c) The AHC of Nb$_{2}$FeB$_{2}$ as a function of energy. }\label{fig:5}
\end{figure}

The intrinsic AHC can be estimated by the Kubo formula (Eq.~(\ref{eq:AHC})) which amounts to the integration of Berry curvature over the occupied states in momentum space, 
\begin{eqnarray}
	{\sigma_{zx}} = -\frac{2\pi e^{2}}{h} \sum_{n} \left[\int_{\rm O}{\frac{d^3 \pmb{k}}{(2\pi)^{3}}}f(\pmb{k}){\Omega_{y}(n,\pmb{k})}\right].
	\label{eq:AHC}	
\end{eqnarray}
Here $\int_O$ stands for the integration of the occupied states.  The Berry curvature distributions in the $k_{y}$=0 plane for Nb$_{2}$FeB$_{2}$ and Ta$_{2}$FeB$_{2}$ are plotted in Fig.\ref{fig:5}(a) and (b), respectively. It is obvious that the positive part and the negative part of Berry curvature cannot cancel each other. According to formula (Eq.~(\ref{eq:AHC})), the AHC $\sigma_{xz}$  of Nb$_{2}$FeB$_{2}$ and Ta$_{2}$FeB$_{2}$ are nonzero. The numerical results for AHC of Nb$_{2}$FeB$_{2}$ and Ta$_{2}$FeB$_{2}$ are $-100 \Omega^{-1}$ cm$^{-1}$, and $-54 \Omega^{-1}$ cm$^{-1}$, respectively. Moreover, we also computed the AHC of Nb$_{2}$FeB$_{2}$ around the Fermi level to mimic the charge doping effect. As shown in Fig.\ref{fig:5}(c), the AHC of Nb$_{2}$FeB$_{2}$ can be enhanced by doping holes.

As we have shown, symmetry is very important for the realizing of AHC in the collinear antiferromagnetism. The symmetry analysis shows that the AHE in the collinear antiferromagnetism without the 'prohibiting' symmetries $\{\mathcal T| \tau\}$ and $\mathcal I \mathcal T$ can be realized in such high-symmetry crystals that do not have two orthogonal mirrors, $C_{2}$ symmetries perpendicular to each other, $C_{4z} \mathcal T$ symmetry, and $C_{6z} \mathcal T$ symmetry. To be specific, if collinear AFM materials have type-I magnetic space group symmetry and the corresponding magnetic point group symmetry is $C_{4}$, $C_{4h}$,$C_{3h}$, $C_{6}$, or $C_{6h}$, these materials have non-zero AHC $\sigma_{xy}$ but zero $\sigma_{xz}$ and $\sigma_{yz}$. On the other hand, if collinear AFM materials have type-I magnetic space group symmetry and the corresponding magnetic point group symmetry is $C_{3}$ or $C_{3i}$, the corresponding AHC $\sigma_{xy}$, $\sigma_{xy}$ and $\sigma_{xy}$ are all non-zero (the high-symmetry axis is along the $z$ direction).
  
{\it Discussion.} Based on the above analysis, there are fourfold advantages for Nb$_{2}$FeB$_{2}$ and Ta$_{2}$FeB$_{2}$ as one class of collinear AFM materials with intrinsic AHC. First, the Nb$_{2}$FeB$_{2}$ and Ta$_{2}$FeB$_{2}$ may be the first class of materials with intrinsic AHC induced by collinear antiferromagnetism. Second, the AHC of Nb$_{2}$FeB$_{2}$ is very large, being the same order of magnitude as those of ferromagnetic and non-collinear AFM metals. Third, the Néel temperatures of both Nb$_{2}$FeB$_{2}$ and Ta$_{2}$FeB$_{2}$ are beyond room temperature, thus the AHE in collinear antiferromagnetism may be realized in room temperature. Fourth, the Nb$_{2}$FeB$_{2}$ and Ta$_{2}$FeB$_{2}$ had been synthesized\cite{NbFeB-1968, TaFeB-1971}, which is very helpful to studying the AHE in collinear antiferromagnetism and exotic physical properties of Nb$_{2}$FeB$_{2}$ and Ta$_{2}$FeB$_{2}$ experimentally.

In summary, we have predicted that two existing collinear antiferromagnetic metals, Nb$_{2}$FeB$_{2}$ and Ta$_{2}$FeB$_{2}$, may have strong AHE above room temperature. Especially, the AHC of Nb$_{2}$FeB$_{2}$ is around $-100  \Omega^{-1}$ cm$^{-1}$, which is the same order of magnitude as those of ferromagnetic metals. Meanwhile, we complete the symmetry groups allowing nonzero AHE in collinear antiferromagnetism. The symmetry principle is of great significance in the study of AHE in collinear antiferromagnetism.

{\it Acknowledgments.} We thank B.-C. Gong for valueable discussions. This work was financially supported by the NSF of China (No.12204533, No. 11934020, No. 11974421, and No. 12134020) and by the National Key R$\&$D Program of China (No. 2019YFA0308603).

%

\begin{thebibliography}{34}
\bibitem [{Nex()}]{Next}%
	\BibitemOpen
	\href@noop {} {\bibinfo  {journal} {Since the present work is focused on the
			AHC in collinear antiferromagnetism, the topological properties of
			Nb$_{2}$FeB$_{2}$ and Ta$_{2}$FeB$_{2}$ will be studied in our next work}\
	}\BibitemShut {NoStop}%
\end{thebibliography}

\end{document}